\documentclass{amsart}

\usepackage{amsmath}    
\usepackage{amssymb}    
\usepackage{dsfont} 	
\usepackage{bm}         
\usepackage{mathtools}  
\usepackage[hypertexnames=false]{hyperref} 
\usepackage[scientific-notation=true]{siunitx}
\usepackage{enumerate}
\usepackage{centernot}
\usepackage{pdflscape}
\usepackage{paralist}
\usepackage{stackrel}
\usepackage[english]{babel}
\usepackage[table]{xcolor}
\usepackage{setspace}

\usepackage[numbers]{natbib}

\usepackage{listings}
\usepackage{color}

\newtheorem*{heuristic*}{Heuristic Definition}
\theoremstyle{definition}
\newcommand{\vet}[1]{{\ensuremath{\mbox{\boldmath $#1$}}}}

\makeatletter
\newcommand{\addresseshere}{%
  \enddoc@text\let\enddoc@text\relax
}

\begin{document}

\title[A Note on a Simple and Practical Randomized Response Framework]
{A Note on a Simple and Practical Randomized Response Framework for Eliciting Sensitive Dichotomous \& Quantitative Information}
\author[C.F.W.\ Peeters]{Carel F.W.\ Peeters}
\address[Carel F.W.\ Peeters]{
Dept.\ of Epidemiology \& Biostatistics \\
Amsterdam Public Health research institute \\
VU University medical center Amsterdam \\
Amsterdam \\
The Netherlands}
\email{cf.peeters@vumc.nl}

\author[G.J.L.M. Lensvelt-Mulders]{Gerty J.L.M.\ Lensvelt-Mulders}
\address[Gerty J.L.M. Lensvelt-Mulders]{
Theory of Sciences and Research Methodology \\
University of Humanistic Studies \\
Utrecht \\
The Netherlands}
\email{g.lensvelt@uvh.nl}

\author[K.\ Lasthuizen]{Karin Lasthuizen}
\address[Karin Lasthuizen]{
School of Management \\
Victoria Business School \\
Victoria University of Wellington \\
Wellington \\
New Zealand}
\email{karin.lasthuizen@vuw.ac.nz}

\begin{abstract}\label{abstract}
\noindent Many issues of interest to social scientists and
policymakers are of a sensitive nature in the sense that they are
intrusive, stigmatizing or incriminating to the respondent. This
results in refusals to cooperate or evasive cooperation in
studies using self-reports. In a seminal article Warner \citep{Warner65}
proposed to curb this problem by generating an artificial
variability in responses to inoculate the individual meaning of
answers to sensitive questions. This procedure was further developed
and extended, and came to be known as the randomized response (RR)
technique. Here, we propose a unified treatment for eliciting
sensitive binary as well as quantitative information with RR based
on a model where the inoculating elements are provided for by the
randomization device. The procedure is simple and we will argue that
its implementation in a computer-assisted setting may have superior
practical capabilities.

\end{abstract}

\maketitle


\section{Introduction}\label{INTRO}
Many issues that are of interest to social scientists and
policymakers are of a sensitive nature. A topic can be said to be
`sensitive' when the disclosure of information regarding this topic
poses to be threatening for the respondent in the form of being
potentially stigmatizing, incriminating or severely intrusive \cite{Lee93,LR90}.
Due to the potential threats, people
have self-representational concerns, leading to refusals to
cooperate or evasive cooperation \citep[see, e.g.,][]{Jung90}.
The research problems associated with these behaviors tend to
inhibit their adequate measurement, and lead one to dismiss the
general assumption of sampling theory that data collected on units
in the sample are accurate representations of the values associated
with the units sampled.

In a seminal article Stanley Warner \citep{Warner65} proposed to use the
element of chance stemming from a randomization device to inoculate
individual responses to sensitive inquiries in order to reduce
nonsampling bias. Subsequently the population estimates of
$\pi_{x}$, the population proportion indulging in sensitive trait X,
will become more accurate. This initial idea grew out to a family of techniques commonly
referred to as \emph{randomized response} (RR), which as a core
characteristic uses the insertion of random error by an element of
chance to provide a respondent optimal privacy protection. For an
in-depth review of developments in RR methods we confine by
referring to Deffaa \citep{Def82}, Fox \& Tracy \citep{FT86}, Chaudhuri \&
Mukerjee \citep{CM88}, and Tracy \& Mangat \citep{DM96}.

The psychological premises of the RR
method should be clearly understood: If the respondent understands
that RR objectively guarantees privacy or trusts the method to
provide full anonymity in the disclosure of information (whether
fully understanding it or not), he or she is relieved from
self-representational concerns and will be more inclined to
cooperate and will do so in a nonevasive manner. While generally
RR is thus thought able to eliminate or relieve both cooperation refusal and
evasive cooperation, we view the RR method mainly as a technique
for relieving the latter. As one has to actually endeavor on
responding to an RR survey before fully grasping its relative merit
and assurances with regard to anonymity and data confidentiality,
it can be expected that RR will actually have a stronger stance
in reducing evasiveness than in reducing response refusal. It is in
this light that the developments in the remainder should be read.

Methods for quantitative RR have been given less attention in the
past than its binary family members and existing models and proposed
randomization efforts often prove impractical. Many new developments
in RR seem absorbed by the technical fix, neglecting its purpose as
a technique for improving the validity of observational data. This
short research note has two aims with regard to communicating our
current work on the RR method: (1) to propose a simple unified
framework for eliciting both binary and quantitative information
with RR (Section \ref{FRAME}) and (2) to describe features of the
incorporation of the proposed RR procedures in a computer-assisted
environment (Section \ref{RAND}). Our framework stresses RR as a
survey technique that should have high practicability as we will
argue in the concluding discussion.

\section{Proposed Framework}\label{FRAME}
An efficient procedure for RR is to let the randomization device
provide for the inoculating response. Richard Morton \citep[as cited in][]{GASH69}
and Robert Boruch \cite{Bor71,Bor72} almost
simultaneously proposed an equivalent dichotomous procedure that has
this specific design characteristic, now commonly referred to as
\emph{forced randomized response} (FRR). Letting the randomization device
provide for the inoculating response combines efficiency,
simplicity, and practicability when developing inquiries into large
numbers of sensitive behaviors. Here we will give an account of the
dichotomous FRR procedure and subsequently extend this model to a
polychotomous appreciation of quantitative RR, so as to come to
certain unity in the RR framework.

\subsection{Binary Forced Randomized Response Model}\label{bin}
Consider a sensitive trait X for which the population is
dichotomous. A random sample of $i = 1, ..., i, ... \,n$ persons is
drawn. The goal is the estimation of the population prevalence of X
($\pi_{x}$). To this purpose each person in the sample is asked the
question: ``Do you possess trait X?" Each respondent receives a
randomization device. In previous studies a pair of dice was used
accompanied with the instructions to answer truthfully if the
outcome is 5, 6, 7, 8, 9 or 10; to answer with `yes' if the outcome
is 2, 3 or 4; or to answer with `no' if the outcome is 11 or 12 \cite{HGBH00}.
We propose to use a (digital)
binary spinner along with the following instructions: Turn the
spinner; if it stops on an empty area, respond with `yes' or `no'
truthfully; if it stops on an area imprinted with `yes', answer
`yes'; if it stops on an area imprinted with `no', answer `no'.

\bigskip
\noindent In this setup we define (using notation by Lang \cite{Lang04})
\begin{align*}
X_{i}&= \left\{\begin{array}{ll}
                 1 & \mbox{\emph{i}th respondent posessing trait
                 X}\\
                 0 & \mbox{otherwise}
                \end{array}
         \right.,\\
W_{i}&= \left\{\begin{array}{lll}
                 1 & \mbox{\emph{i}th respondent randomly requested to respond truthfully}\\
                 2 & \mbox{\emph{i}th respondent randomly directed to `yes' response}\\
                 3 & \mbox{\emph{i}th respondent randomly directed to `no' response}
                \end{array}
         \right.,\\
Z_{i}&= \left\{\begin{array}{ll}
                 1 & \mbox{\emph{i}th respondent answering question with `yes'}\\
                 0 & \mbox{\emph{i}th respondent answering question with `no'}
                \end{array}
         \right.,
\end{align*}
and
\begin{align*}
p_{1}&= \mbox{probability of a request on a truthful answer}\\
p_{2}&= \mbox{probability of a forced `yes'}\\
p_{3}&= \mbox{probability of a forced `no'}
\end{align*}
with $p_{1}+p_{2}+p_{3}=1$.

Taking into account the assumptions regarding RR, events can be
expressed in probabilities, so that
\begin{align}\label{popprop}
\nonumber P(Z_{i}=1)=\lambda = &~P(Z_{i}=1\mid W_{i}=1,
X_{i}=1)\cdot
P(W_{i}=1)\cdot P(X_{i}=1) \,+ \\\nonumber
&~P(Z_{i}=1\mid W_{i}=2)\cdot P(W_{i}=2)\\
= &~p_{1}\pi_{x}+p_{2}.
\end{align}
As the probability of a `yes' response can be estimated by the
sample proportion answering `yes' $(\hat{\lambda}=n^{-1}\sum_{i}Z_{i})$ and as $p_{1}$
and $p_{2}$ are fixed by design, an unbiased moment estimator of the
population prevalence of X can be obtained by
\begin{equation}
\hat{\pi}_{x}=\frac{\hat{\lambda}-p_{2}}{p_{1}},
\end{equation}
with sampling variance
\begin{equation}
Var(\hat{\pi}_{x})= \frac{\lambda(1-\lambda)}{np_{1}^{2}},
\end{equation}
and unbiased sampling estimate
\begin{equation}
\widehat{Var}(\hat{\pi}_{x})=\frac{\hat{\lambda}(1-\hat{\lambda})}{(n-1)p_{1}^{2}}.
\end{equation}

\begin{sloppypar}
\subsection{Discrete Quantitative Forced Randomized Response
Model}\label{quant} In inquiries into sensitive behaviors not only
questions of prevalence are of interest, but also questions of
frequency of occurrence. To evaluate frequency of occurrence, RR
models must be used that deal with quantitative data. Here, a
discrete quantitative FRR model is developed for which previous work
by Greenberg \emph{et al.} \cite{GASH71}, Eriksson \cite{Erik73}, Liu \& Chow
\cite{LC76}, and Stem \& Steinhorst \cite{SS84} is acknowledged. Consider a
sensitive trait X, for which the population is supposed to be
continuous or discretely quantitative. A random sample of $i = 1,
..., i, ... \,n$ persons is drawn. Instead of focussing on a mean we
can device a model which redirects X to be discrete, assuming values
$x_{1}, ..., x_{k}$ with respective unknown true proportions
$\pi_{1}, ..., \pi_{k}$, where $0\leq\pi_{j}\leq1$ $(j=1, ..., k)$
and $\sum_{j}\pi_{j}=1$. Each of those values can be assigned a
category of numbers, for example in a setup where k = 6: The
responses 1, 2, 3, 4, 5 and 6 could correspond with the
respective categories `0', `1 time', `2 to 3 times', `4 to 5 times',
`6 to 10 times', and `more than 10 times'. We are then interested in
estimating the proportion in each of $k$ discrete quantitative
categories, providing for a multi-proportional or polychotomous
appreciation of quantitative RR \cite{CFWP05}.
\end{sloppypar}

For such a setup we could again use two dice as randomizing device.
But in this case 2 throws with 3 dice are needed, where the
selection probabilities of $p_{2}$ and $p_{3}$ (as in the previous
section) require for the third die to be thrown. The number that
turns up in this throw gives the forced response. There are two
problems associated with the use of three dice as randomization
device. First, two-step procedures introduce an
extra source of error due to increased respondent burden. Second,
using dice provides a setup where any piece of quantitative
information can be misclassified into exactly 6 discrete categories.
To move beyond these 6 categories and to decrease the respondent
burden one could also supply each respondent with a (digital)
discrete quantitative spinner such as given in Section \ref{RAND}
along with the following instructions: Turn the spinner; if it stops
on an empty area, respond with `1', `2', ..., or `k' truthfully; if
it stops on an area imprinted with `1', `2', ..., or `k', respond
accordingly. The selection probabilities for the truthful answer and
the forced responses are $p$ and $p_{j}$ respectively, with $j = 1,
..., k$ and $\sum_{j}p_{j}=1-p$.

In this setup we define
\begin{align*}
X_{i}&= \left\{\begin{array}{lll}
                 1 & \mbox{\emph{i}th respondent indulging category 1 times in X}\\
                 \vdots & \mbox{\vdots}\\
                 k & \mbox{\emph{i}th respondent indulging category \emph{k} times in X}
                \end{array}
         \right.,\\
W_{i}&= \left\{\begin{array}{lll}
                 1 & \mbox{\emph{i}th respondent randomly requested to answer truthfully}\\
                 \vdots & \mbox{\vdots}\\
                 k+1 & \mbox{\emph{i}th respondent randomly requested to respond with category \emph{k}}
                \end{array}
         \right.,\\
Z_{i}&= \left\{\begin{array}{lll}
                 1 & \mbox{\emph{i}th respondent answering question with category 1}\\
                 \vdots & \mbox{\vdots}\\
                 k & \mbox{\emph{i}th respondent answering question with category \emph{k}}
                \end{array}
         \right.,
\end{align*}
and
\begin{align*}
p&= \mbox{probability of a request on a truthful answer}\\
p_{1}&= \mbox{probability of a forced category 1 answer}\\
\vdots& \mbox{}\\
p_{k}&= \mbox{probability of a forced category \emph{k} answer}
\end{align*}
with $p+\sum_{j}p_{j}=1$.

If we let $\lambda_{j}$ denote the probability of an affirmative
quantitative response, again with $j$ denoting the response pointing
to one of our $k$ discrete quantitative categories and taking into
account the assumptions regarding RR, it can then be shown analogous
to (\ref{popprop}) that
\begin{align}\label{popprop2}
\nonumber P(Z_{i}=j)=\lambda_{j} = &~P(Z_{i}=j\mid W_{i}=1,
X_{i}=j)\cdot
P(W_{i}=1)\cdot P(X_{i}=j) ~+\\
\nonumber &~P(Z_{i}=j\mid W_{i}=j+1)\cdot P(W_{i}=j+1)\\
= &~p\pi_{j}+p_{j}.
\end{align}
As the probability of a certain quantitative response within our $k$
category parameter space can be estimated by the sample proportion
giving an affirmative response to that certain quantitative
category $(\hat{\lambda}_{j}=n^{-1}\sum_{i}[Z_{i}=j]$, where $[\cdot]$ denotes the characteristic function$)$, and as $p$ and $p_{j}$ are fixed by design, each of our
$k$ population proportions can be unbiasedly estimated by solving
(\ref{popprop2}) for $\pi_{j}$:
\begin{equation}
\hat{\pi}_{j}=\frac{\hat{\lambda}_{j}-p_{j}}{p}.
\end{equation}
The property of unbiasedness follows from
\begin{eqnarray*}
\nonumber E(\hat{\pi}_{j}) = \frac{1}{p}\left(E(\hat{\lambda}_{j})-p_{j}\right)
= \frac{1}{p}(p\pi_{j}+p_{j}-p_{j})
= \pi_{j}.
\end{eqnarray*}
Utilizing the quality of a multinomial distribution as a joint
structure of binomials, the sampling variance for $\hat{\pi}_{j}$ is
\begin{eqnarray}
Var(\hat{\pi}_{j}) = \frac{1}{p^{2}}Var(\hat{\lambda}_{j})
=\frac{1}{p^{2}}\frac{1}{n}(p\pi_{j}+p_{j})(1-p\pi_{j}-p_{j})
= \frac{\lambda_{j}(1-\lambda_{j})}{np^{2}},
\end{eqnarray}
with unbiased sampling estimate
\begin{equation}
\widehat{Var}(\hat{\pi}_{j})=\frac{\hat{\lambda}_{j}(1-\hat{\lambda}_{j})}{(n-1)p^{2}}.
\end{equation}

\subsection{General Form}\label{common}
Let $\vet{\lambda}=(\lambda_{1}, ..., \lambda_{k})^{t}$ denote the
probabilities of the observed responses with $1, ..., k$ categories,
and let $\vet{\pi}=(\pi_{1}, ..., \pi_{k})^{t}$ denote the
probabilities of true answers/status with $1, ..., k$ categories.
The models as developed here can then be generally captured in the
function
\begin{equation}
\vet{\lambda}=\vet{P\pi},
\end{equation}
so that the moment estimator is given by \cite{CM88}
\begin{equation}\label{estpi}
\vet{\hat{\pi}}=\vet{P^{-1}\hat{\lambda}}.
\end{equation}
In the aforementioned
\begin{equation}
\vet{P}=({P}_{j^{o},j^{t}})_{k\times k} =
  \left[\begin{array}{ccc}
  P_{11} & \cdots & P_{1k}\\
  \vdots & \ddots & \vdots\\
  P_{k1} & \cdots & P_{kk}
  \end{array} \right],
  \end{equation}
the $k\times k$ matrix of conditional misclassification
probabilities where $j^{o}$ denotes observed answers and $j^{t}$
denotes true status. Thus, when having a nonsingular $\vet{P}$ and
an unbiased estimate $\vet{\hat{\lambda}}$ of $\vet{\lambda}$, then
$\vet{\pi}$ can be estimated by (\ref{estpi}). As van den Hout and
van der Heijden \cite{HH04} point out, the assumption that $\vet{P}$ is
nonsingular does not impose much restriction on the
misclassification design matrix as $\vet{P}^{-1}$ exists when the
diagonal of $\vet{P}$ dominates, meaning: $P_{j^{o},j^{t}} > 1/2$,
for $j^{o}=j^{t}$. This is reasonable as the diagonal elements
represent correct classification and these probabilities should be
taken relatively high for the design to be efficient.

\section{Computer-Assisted Randomized Response}\label{RAND}
The incorporation of computer-assisted self interviewing (CASI) and
RR in computer-assisted randomized response (CARR) interviewing has
been proposed and successfully tested by Musch, Br\"{o}der \&
Klauer \cite{MBK01}, Lensvelt-Mulders \emph{et al.}
\cite{LHLG06} and Lensvelt-Mulders \& Boeije \cite{LB07}. CARR could prove to be
very important for future inquiries into socially sensitive issues
as other computer-based techniques lose their magic due to
penetration of mainstream culture \cite{Boeije2002,LHLG06}, as
respondents' awareness of the negative possibilities of computers
regarding his or her privacy issues increases with increasing
`computer-literacy', and as the use of internet makes possible the
dissemination of data on a large scale which may possibly relieve
the efficiency problems inherent in RR designs \cite{MBK01}.
Moreover, Boeije \& Lensvelt-Mulders \cite{Boeije2002} and Lensvelt-Mulders \&
Boeije \cite{LB07} have shown that respondents prefer computer-assisted RR above
paper-and-pencil and face-to-face RR.

The use of CARR makes possible the use of accompanying digital
randomizers. Here we propose digital spinners for binary and
multi-proportional FRR. Spinners have the advantage of being widely
known at least across western populations and it may be argued that, when unfamiliar
with spinners, their concept may be easily grasped. Also, two important problems are
solved concerning the randomizers when a digital make-up is used.
The first of which is randomness. Randomness has proven hard to
achieve with physical spinners. For example, Stem and Steinhorst
\cite{SS84} used a physical spinner in a mail questionnaire. This spinner
could get bend in the mail, and subsequently affect the randomness
of the device. With sufficient programming the digital randomizers
give very acceptable and consistent levels of pseudo-randomness. The
second problem alleviated with a digital make-up of the randomizing
device is the problem of inconclusive outcomes. In the same study by
Stem and Steinhorst, an answer could land on a line on their
physical spinner, proving the need for additional instructions,
which increases the cognitive load for the respondent. Sufficient
programming curbs this problem.

\begin{figure}[tp] \centering
\resizebox{12cm}{!}{\includegraphics{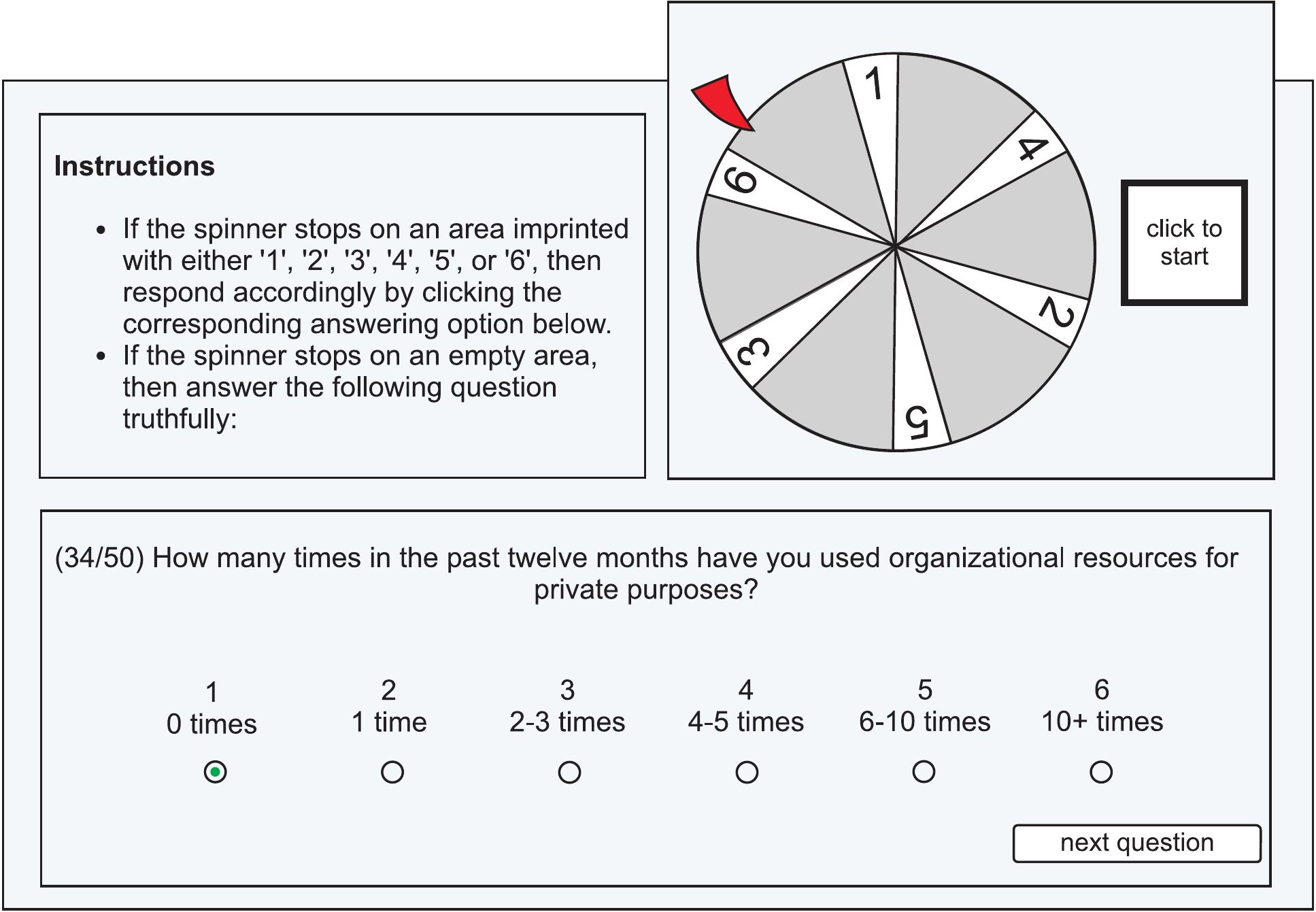}} \caption{A
Computer-Assisted Forced RR Questionnaire}\label{QUESTIFig}
\end{figure}

Figure 1\footnote{The spinners were developed by Peeters \cite{CFWP05} and
programmed by Rens Lensvelt of WetenschapWerkt. Web-page
questionnaires for CASI and CARR as used in Peeters \cite{CFWP05} are
properties of C.F.W. Peeters, WetenschapWerkt, Utrecht University,
and VU University Amsterdam.} gives an impression of our
incorporation of a (discrete quantitative) FRR spinner into CASI to
make for a CARR questionnaire. The upper left corner contains
instructions as in Section \ref{FRAME}, the upper right corner
contains the randomizer, and the bottom part contains the sensitive
question with accompanying answering categories. The discrete
quantitative FRR spinner of Figure \ref{QUESTIFig} is constructed with $k=6$, a
$p$ of three-fourths and a $\sum_{j}p_{j}$ of one-fourth. These selection
probabilities are translated into the randomization device by
converting probabilities into degrees. Moriarty and Wiseman \cite{MW76}
give evidence that respondents, due to the involved permutations
leading to a certain outcome, misperceive selection probabilities
when dice are used as a randomization device. If the misperceiving
of probabilities can be incorporated in the randomization device, it
is possible to provide for sufficient protection, while pertaining
relatively high efficiency. To enhance the misperceiving of
selection probabilities the empty and imprinted areas are evenly
divided over the spinner, which is divided in 24 sub-areas. This
gives every empty area a selection probability of 3/24 ($6\cdot3/24
= 3/4$) and the analogous selection probability of each of the 6
forced numerical responses is 1/24. These probabilities and number
of categories are by no way necessary. The selection probabilities
and number of discrete quantitative categories can be chosen to
convenience.

Note that in the aforementioned setup of the discrete quantitative FRR
spinner a one-step solution is given for the multiple-step
misclassification of sensitive data in 6 categories using 3 dice as
described in Section \ref{quant}. Also note that by replacing the
numbers in the imprinted areas with `yes' and `no', a binary
computer-assisted FRR questionnaire is obtained.

\section{Discussion}\label{DISCUSS}
\begin{sloppypar}
The unified framework for binary and quantitative RR provides
several advantages. The design can be adjusted to any efficiency
level by selecting certain parameters of the randomization device.
Additionally, the spinners are easily incorporated in computerized settings and
provide the opportunity of utilizing the psychological advantage of
the misperceiving of actual misclassification probabilities \cite{FT86,CFWP05}.

Furthermore, sensitive quantitative information can be misclassified
in $k$ discrete categories, where the scope of the frequencies
denoted with each category can be adapted to the researcher's need.
In previous quantitative designs the desired population means
regarding certain sensitive behaviors had to be estimated before
any field research efforts, so as to adapt the setup of the
randomizer to the expected range of frequencies in the population.
After all, the range of responses to the innocuous question has to
be similar to the range of possible responses to the sensitive
question if one is to unbiasedly estimate frequency of occurrence
while still providing sufficient respondent protection \cite[see, e.g.,][]{LC76,SS84}. 
In the proposed unified approach, the categories can be very easily adopted
to expected frequencies while retaining setup and efficiency, as the
forced responses have a range that is equal to the range of possible
responses on the sensitive question, due to the multi-proportional
construction. Respondent protection is thus also statistically
guaranteed in the quantitative RR setup.

Notwithstanding the relieve RR provides with regard to self-representational concerns, 
respondents may still cheat, that is, they may not comply with the RR instructions. 
Assuming clear instructions and respondent understanding, two constructs are theorized 
to be responsible for non-compliance: respondent jeopardy and risk of suspicion \cite{GASH77}. 
The former refers to guilty respondents' risk of being identified as such when responding 
affirmatively. The latter refers to innocent respondents' risk of being identified as 
guilty when responding affirmatively. These risks may advocate self-protective response 
behavior: evasive answering irrespective of the outcome of the randomization device. 
Provisions exist in the form of parameter selection and design symmetry.

The optimization of parameter selection for both the binary and quantitative FRR models in terms 
of reconciling respondent hazards and efficiency can be captured in the Bayesian framework 
for conditional RR probabilities proposed by Lanke \cite{Lanke75,Lanke76} and Greenberg \emph{et al.} \cite{GASH77}. 
Moreover, the proposed FRR design is symmetric \cite{BD76}, meaning that no possible 
response option, in itself, conveys information on one's true status. While less efficient than 
asymmetric designs, there is evidence that symmetric designs spur less cheating due to reduced 
risk of suspicion relative to their asymmetric counterparts \cite{OMZM09}.

These provisions may not completely eliminate self-protective response behavior. 
It may subsequently be possible that the percentage of affirmative responses 
(in certain response categories) falls below chance level. We would then have 
an estimator that lies outside the interior of the parameter space, implying 
that the moment estimator (10) is no longer equivalent to the maximum likelihood 
estimator. It is here that we may also see the analytical advantage of the unified 
RR framework in that it allows for a certain unity in the analysis of dichotomous 
and quantitative RR data. van Den Hout and Van Der Heijden \cite{HH04} give an elegant 
framework for analyzing RR data based on a log-linear latent class model analogy. 
Their framework unifies the chi-square test of independence and numerical maximum 
likelihood estimation for models of the general form (9). Their log-linear RR 
approach may be extended to provide prevalence estimates corrected for self-protective 
response behavior \cite{CHHB07}.

Adding to the attractiveness of practicality, one may note that it is straightforward 
to incorporate Likert-scale type questions (e.g., to study sensitive attitudes) 
into the quantitative FRR setup, as the distinct categories 1-6 could be
labeled as a Likert scale $(1 = \mbox{never}, ..., 6 =
\mbox{always})$. The RR framework given earlier thus provides for simple binary and
quantitative RR models whose incorporation in a computerized setting
may prove to be more practically feasible in real research settings.
The models and CARR questionnaires as described above have been
tested in our lab and field-research efforts are currently being
undertaken.
\end{sloppypar}

\section*{Acknowledgements}
At the time of writing C.F.W.P. was affiliated with the Department of
Methodology \& Statistics at Utrecht University and K.L. was affiliated 
with the Department of Governance Studies, VU University Amsterdam.
This version is a postprint of:
Peeters, C.F.W., Lensvelt-Mulders, G.J.L.M., \& Lasthuizen, K. (2010). 
A Note on a Simple and Practical Randomized Response Framework for Eliciting Sensitive Dichotomous and Quantitative Information.
\emph{Sociological Methods \& Research}, 39: 283--296.



\vspace{1cm}
\addresseshere

\end{document}